# Initial electron thermalization in metals measured by attosecond transient absorption spectroscopy

Bethany R. de Roulet,[1] Lorenz Drescher,[1,2] Shunsuke A. Sato,[3,4] and Stephen R. Leone[1,2,5]

[1]Department of Chemistry, University of California, Berkeley, CA 94720
[2]Department of Physics, University of California, Berkeley, CA 94720
[3]Center for Computational Sciences, University of Tsukuba, Tsukuba, Ibaraki 305-8577, Japan
[4]Max Planck Institute for the Structure and Dynamics of Matter, Luruper Chaussee 149, 22761 Hamburg, Germany
[5]Chemical Sciences Division, Lawrence Berkeley National Laboratory, Berkeley, CA 94720

## Abstract

Understanding initial electron thermalization has relevance to both fundamental scientific knowledge and application to the construction of novel devices. In this study, attosecond transient absorption is used to directly measure initial electron thermalization times of $38 \pm 8$ fs, $15 \pm 3$ fs, $4.2 \pm 1$ fs, and $2.0 \pm 0.3$ fs for Mg, Pt, Fe, and Co, respectively. Through time dependent density function theory calculations, it is shown that the fast electron thermalization observed in Fe and Co is correlated with a strong local field effect. We find that a simple analytical model can be used to calculate the initial electron thermalization time measured by the transient extreme ultraviolet absorption spectroscopy method performed here. Our results suggest that the most significant contributions to the initial electron thermalization times are the basic metal properties of the density of states volume available for scattering and screened electron interaction. Many-body effects contribute less, but still significantly to the initial electron thermalization time. Ultimately the information gained through this study shows the unique view that attosecond transient absorption spectroscopy contributes to unraveling and monitoring electron dynamics and its connection to many-body effects in metals and beyond.

## I. Introduction

What processes contribute to non-equilibrium electron thermalization has been a long-standing question for many-body physics. The nature of electron thermalization is not only a fundamental question, but it is of interest for device applications. It has been shown that spatial gradients created by non-equilibrium electron distributions result in transient behavior, such as ultrafast magnetic phenomena.[1,2] Additionally, energy transport of non-equilibrium electrons can be up to two orders of magnitude greater than their equilibrium equivalents.[3,4] These unique properties driven by electron thermalization are significant for the development of novel devices ranging from magnetic storage devices to hot carrier solar cells.

In the pursuit of understanding electron thermalization, metals are the natural starting place as they have the simplest material properties and have a large range of well-developed theory available. The photoexcitation process in metals promotes electrons from states below the Fermi level to above it, leaving these photoexcited electrons in a highly non-equilibrium, nascent, distribution. This nascent distribution decays on the femtosecond timescale to a hot-Fermi Dirac distribution through electron-electron collisions.[5] Finally electrons return to equilibrium by transferring their energy to phonons, defects, surfaces, or impurities. Some of the main factors dictating the efficiency and therefore the timescale of the initial electron thermalization, i.e. the transition from the nascent to hot Fermi-Dirac distribution, are the number of states available to facilitate electron-electron scattering, the strength of coupling between scattering partners, and the temperature of the non-excited scattering partners.[6, 7]

From an experimental perspective, very few techniques operate on a fast-enough timescale to directly measure the initial electron thermalization, which can occur within a femtosecond of excitation.[7] Two-photon photoemission (2PPE) is a common spectroscopic technique for measuring the lifetime of an electron in an excited state, however the final state of the probe electron is outside the material, which makes the probe blind to material responses related to electron-electron interactions. 2PPE is also limited to a low excitation density in order to avoid space charging of ejected photoelectrons, and the method is subject to a dependence between measured electron lifetimes and the temporal duration of the excitation pulses employed.[8] Although valuable insight has been gained through 2PPE studies, knowledge of non-equilibrium electron thermalization could greatly benefit from another technique, such as attosecond transient extreme ultraviolet light (XUV) absorption spectroscopy (ATAS), which probes the bulk of the material, is robust to high excitation densities, and has few femtosecond or subfemtosecond time resolution.

Previous ATAS studies in metals have measured the fluence dependence of electron thermalization times in Ni, showed the influence of the local field effect on spectral features and charge localization in Ti, and observed broadening of the electron distribution and a shift of the chemical potential in Al.[9-11] In this work we broaden previous 2PPE and ATAS studies by using ATAS to measure the initial electron thermalization time in four metals (magnesium, platinum, iron, and cobalt) in order to gain insight into the fundamental relationship between the measured initial electron thermalization time, many body effects, and material properties.

This is achieved by first quantifying the static XUV absorption spectra, the changes in XUV absorption upon optical excitation, and the electron thermalization times with ATAS measurements. Next spectral line shape decomposition, provided through the *ab-initio* modeling, is used to unravel the relevant excited state population changes and many-body interactions that occur after photoexcitation in order to correlate these effects with the experimentally measured initial thermalization times.[12] Finally, the measured initial electron thermalization times are compared to thermalization times calculated by a simple analytical model to provide a means of estimating the thermalization time without the need for measurements and to gain insight into the factors that contribute to the thermalization time.

## II. Results and Discussion

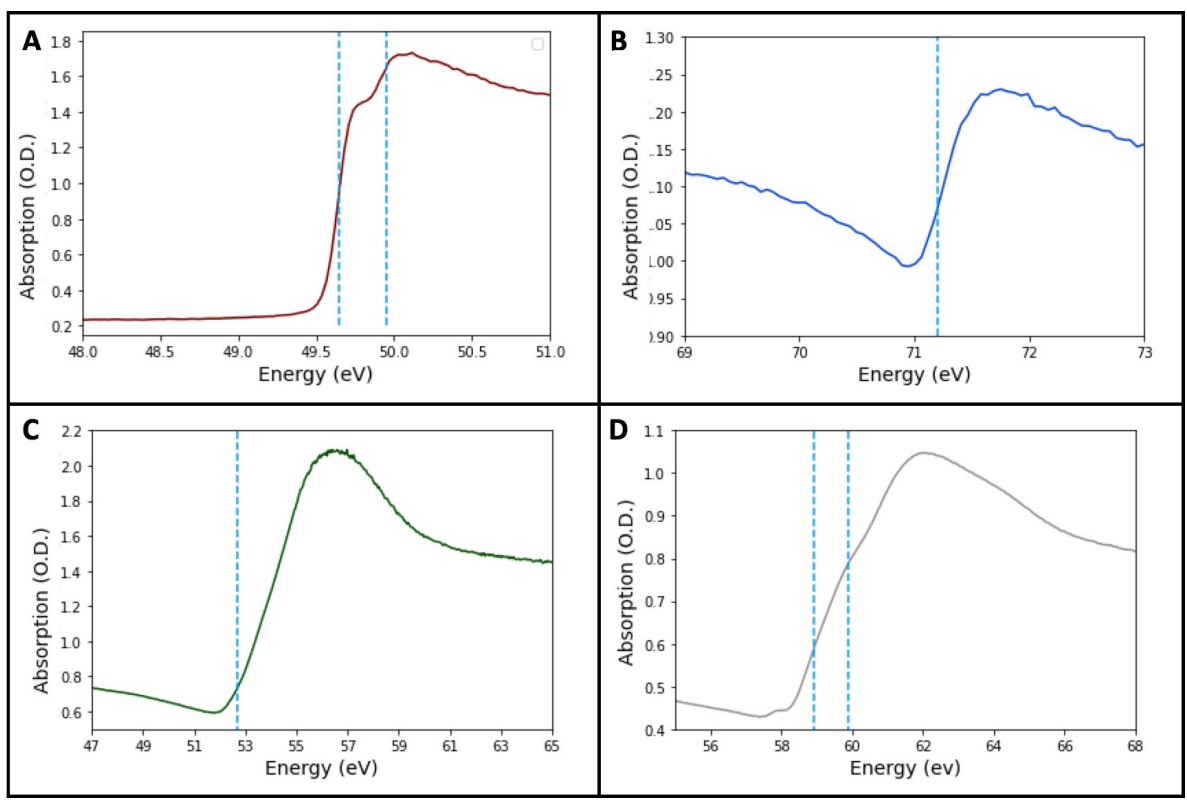

*Fig. 1.* Equilibrium XUV absorption at the Mg $L_{2,3}$ –edges (A), Pt $N_7$ – edge (B), Fe $M_{2,3}$-edges (C) and Co $M_{2,3}$-edges (D). The dashed lines indicate the energies of the expected static absorption edges from previous x-ray absorption measurements (in some cases with a large spin-orbit splitting).[13-16]

The metal films studied here were grown by thermal evaporation onto silicon nitride membranes (Appendix A). The transient XUV absorption apparatus is described in detail in Appendix B. The static XUV absorption spectrum was used to determine film thickness (Fig. 1).

In these measurements, the XUV absorption of the silicon nitride membrane was subtracted from the sample in order to measure the XUV absorption of the metal alone. Using tabulated XUV absorption data for Mg, Fe, and Co, and a calibrated sample for Pt, it was determined that the Mg, Pt, Fe, and Co films were $130 \pm 1$ nm, $35 \pm 1$ nm, $18 \pm 1$ nm, and $9 \pm 2$ nm thick, respectively (Appendix B). The XUV light used in absorption measurements was produced through the process of high harmonic generation (HHG) in argon gas driven by 4 fs, 1.3 eV – 2.6 eV pulses at a 1 kHz repetition rate (Appendix C). The resulting attosecond pulse train of XUV light continuously spanned from 20 eV - 80 eV. In the transient experiments an optical pump with the same bandwidth as the HHG driving field and a fluence of 20 mJ cm$^{-2}$, 23 mJ cm$^{-2}$, 29 mJ cm$^{-2}$, and 21 mJ cm$^{-2}$ was used to excite charge carriers in Mg, Pt, Fe, and Co, respectively. The density of excited charge carriers was $2.7 \times 10^{20}$ cm$^{-3}$, $1.1 \times 10^{21}$ cm$^{-3}$, $1.9 \times 10^{22}$ cm$^{-3}$, and $1.7 \times 10^{22}$ cm$^{-3}$ for Mg, Pt, Fe, and Co, respectively (Appendix D). The resulting change in absorption was then calculated for increasing time delays between the optical pump and XUV absorption probe, where negative time delays correspond to the pump arriving after the probe.

The resulting change in absorption, $dA = -\log \frac{I_{pump\,on}}{I_{pump\,off}}$, was then calculated for increasing time delays between the optical pump and XUV absorption probe. Here, negative time delays correspond to the pump arriving after the probe and positive time delays correspond to the optical pump arriving before the probe. Further detail on the experimental methods can be found in Appendix C.

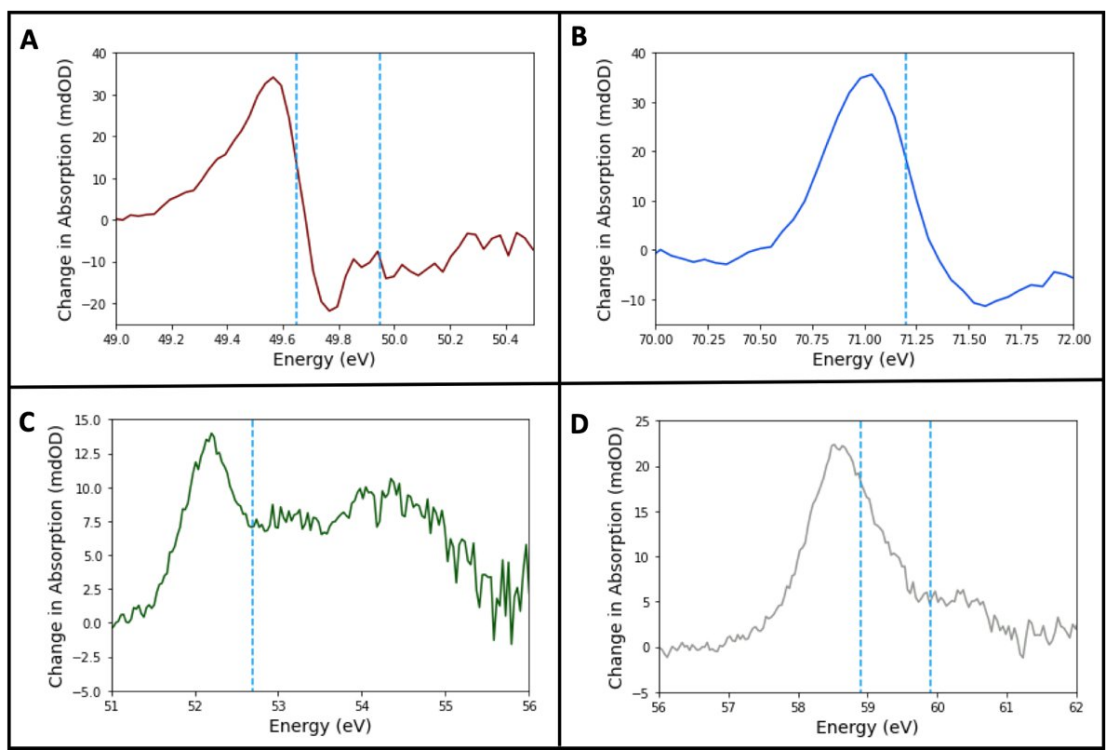

Fig. 2. . Panels A – D show the spectral line shape of the change in for Mg, Pt, Fe, and Co, respectively. The dashed lines correspond to the energy of the static absorption edges.

The transient absorption spectra (Appendix E) were integrated over the time ranges of 60 – 70 fs, 35 – 45 fs, 15 – 25 fs, and 2 – 12 fs for Mg, Pt, Fe, and Co, respectively, to obtain the transient absorption line shapes (Fig. 2, panels A – D). All four metals display a large positive change in absorption at energies below the energy of their respective static absorption edges (Fig. 2 dashed lines). This positive change in absorption has been observed in transient XUV absorption spectra of other metals and has been previously identified as a state filling or state blocking change in the spectra. [9-12,17,18] At energies directly above the Mg $L_3$-edge (49.65 eV) and the Pt $N_7$-edge (71.2 eV) a negative change in absorption is observed (Fig. 2 A, B), while the change in absorption remains positive at energies above the Fe $M_{2,3}$-edge (52.7 eV) and Co $M_3$-edge (58.9 eV) (Fig. 1 C, D). The origin of the spectral features above the static absorption edge energy contain contributions from state filling and many-body effects. Both the state filling / opening contributions and many body effects are discussed in more detail later in section B.

## A. Temporal evolution of the state filling feature

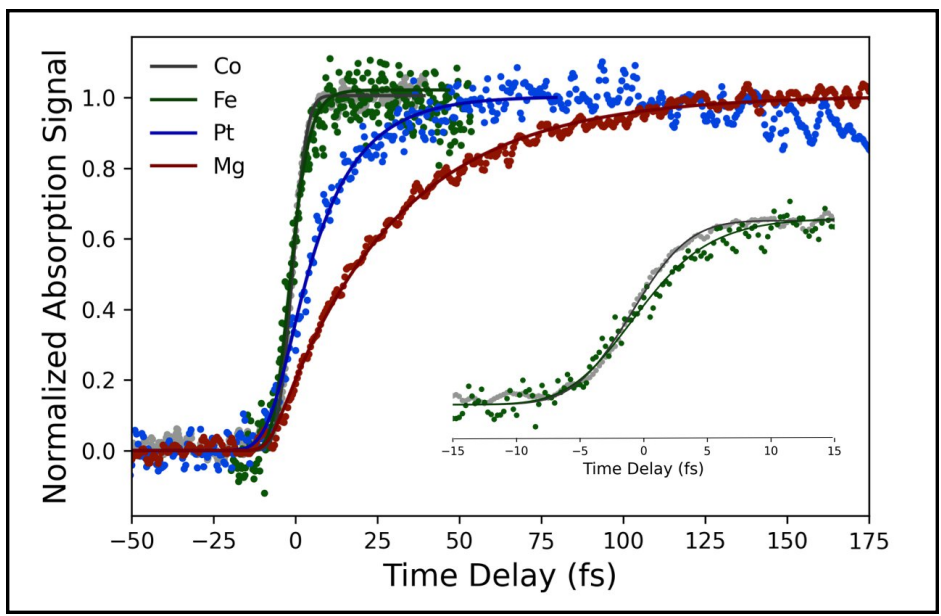

*Fig. 3. Integrating the change in absorption of the positive portion (newly created absorption) of the state-blocking feature (points) shows the dramatic difference in response to optical pumping for Mg (red), Pt (blue), Fe (green), and Co (gray). This is then fit with an exponentially modified Gaussian function to quantify the rise-time of the signal (solid lines). The slight difference in rise time between Fe and Co is highlighted in the inset.*

To show the evolution of the state opening feature in the time domain, the transient absorption spectra were integrated directly below the static absorption edge; 49.2 eV – 49.7 eV, 70.5 eV – 71.3 eV, 51.8 eV - 52.6 eV, and 56.5 eV – 65.2 eV for Mg, Pt, Fe, and Co, respectively (Fig. 3, points). The inset in Fig. 3 compares Co (gray) and Fe (green) from -15 to 15 fs in order to better visualize the difference in the temporal response for these two metals. The experimental data were fit with an exponentially modified Gaussian function (Fig. 3, solid lines) to determine rise times of 38 ± 8 fs, 15 ± 3 fs, 4.2 ± 1 fs, and 2.0 ± 0.3 fs for Mg, Pt, Fe, and Co, respectively. More details on the integration and fitting can be found in Appendix F.

## B. Discerning contributions to the transient absorption spectral lineshape with TDDFT

Determining the factors that contribute to the transient absorption lineshape aid in understanding the underlying processes in the metal that leads to the dramatic difference observed in the timescales of the state opening feature. Figure 4 shows time dependent density functional theory (TDDFT) computation results for three spectral aspects, the state filling / opening component (orange solid line), local field effect (LFE) component (purple solid line), and the change in absorption coefficient (blue solid line) for Mg (Fig. 4A) and Co (Fig. 4B). The full calculated change in absorption coefficient (blue) is to be compared to the experimental transients for Mg (red dashed line) and Co (gray dashed line). More detail on the TDDFT calculations can be found in Appendix G.

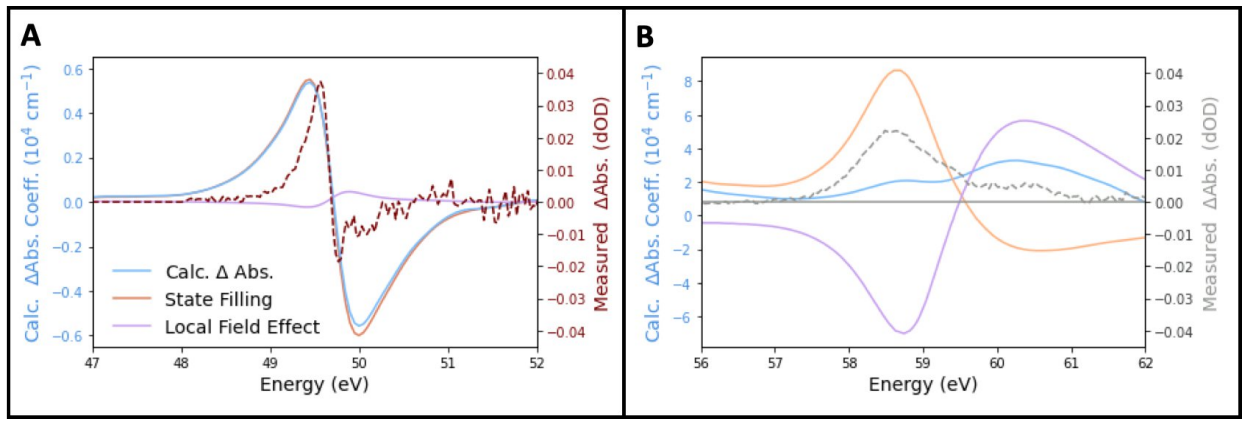

Fig. 4. TDDFT calculations of the full change in absorption coefficient (blue), state filling contribution (orange), and local field effect contribution (purple) show that the measured change in absorption lineshape of the change in absorption signal for Mg (dashed red line) comes primarily from state filling / opening (A), while both state filling / opening and the local field effect contribute to the measured change in absorption lineshape of Co (dashed gray line) (B).

The positive feature of the state filling / opening contribution refers to an observation of a non-equilibrium distribution of holes and electrons in the XUV absorption probing. When the optical pump promotes electrons from below to above the Fermi level new transitions are opened to the XUV absorption probe below the static XUV absorption edge and previously available transitions above the static XUV absorption edge are blocked. This is viewed as a characteristic increase followed by a decrease in the change of absorption lineshape versus photon energy, centered around the static XUV absorption edge.

The LFE is a many-body effect, where a charge carrier is influenced by the electric field created by other nearby charges.[19-21] Here, charge carriers excited by the optical pump couple with other excited carriers, resulting in a collective excitation mode. The LFE shifts the dielectric function to higher energy, thereby moving the XUV absorption edge to higher energy.[22-25] This is seen as a negative followed by a positive change in absorption versus photon energy centered around the static XUV absorption edge.

The experimentally measured transient absorption lineshape of Mg is well-described by the calculated state-filling / opening feature (Fig. 4A), while the experimentally measured lineshape of Co includes contributions of both state filling / opening and the LFE (Fig. 4B). The LFE in Co reduces the intensity of the positive change in absorption attributed to state opening and the negative portion of the state filling contribution is completely masked by the LFE contribution. The change in absorption lineshape of Mg and Co is reasonably well matched by the TDDFT calculations, however there are two notable differences between the full TDDFT calculations and measurements. First, the calculated spectra are broader than the measurements. This possibly stems from the broadening parameter in the TDDFT calculations overestimating the experimental linewidth. The TDDFT calculations may also overestimate the LFE. For Mg this suggests that the small calculated LFE component may be negligible. Co has a much stronger LFE component, so the overestimation in the TDDFT calculations leads to the 60.3 eV peak carrying more spectral weight than the 58.5 eV peak. Despite these two differences, the trend that ATAS measurements of Mg can be viewed primarily as a direct observation of state filling / opening whereas Co includes both state filling and LFE contributions still holds.

      Due to computational time and complexity, the line shape contributions were not calculated for Pt or Fe; the results of Mg and Co can in principle be extended to these metals. The experimental line shape of Pt (Fig. 2B) is similar to Mg, showing that state filling / opening is the major contributing factor for both Pt and Mg. In similar fashion, the lineshape of Fe (Fig. 2C) follows the line shape observed in Co, suggesting that both state filling / opening and LFE contribute to the line shape of these metals. Therefore, it can be concluded that state filling / opening contributes to the transient absorption line shape of all four metals and the LFE additionally contributes to the line shape of Fe and Co.

      The spectral analysis confirms the earlier assertion that changes in the electron occupation (the state opening feature) were temporally quantified in figure 3. Therefore, the timescales reported here correspond to the initial electron thermalization. It also indicates a striking correlation between a faster initial electron thermalization time and a strong LFE. This suggests that intrinsic properties of the metals are responsible for the variety of initial electron thermalization times that are observed.

## C. Calculating the initial electron thermalization time

Previous theoretical work developed for 2PPE measurements on metals correlated the initial electron thermalization time with intrinsic properties of the metals.[6,26,27] From Fermi liquid theory, the thermalization rate of a single excited electron is[23,24, 26,27]

$$\frac{1}{\tau} = A \frac{(\pi k_B T_e)^2 + E^2}{1 + e^{\frac{-E}{k_B T_e}}} \qquad (eq.1)$$

where $\tau$ is the thermalization time of the excited electron, $A$ is the characteristic electron scattering constant, $k_B$ is the Boltzmann constant, $T_e$ is the electron temperature of the electron bath (the non-excited electron scattering partners), and E is the energy of the excited electron relative to the Fermi level (i.e. $E_F$ = 0 eV). Several transient spectroscopy studies of metals have shown that a larger electron temperature (higher excited charge carrier densities) lead to faster electron thermalization for a particular metal.[9,28] The electron temperatures for the experimental conditions used here are 1300 K, 2300 K, 3700 K, and 3400 K for Mg, Pt, Fe, and Co, respectively (Appendix H). However, because $k_B T_e$ << E for all of the metal films eq.1 reduces to $\frac{1}{\tau} \approx A E^2$. Therefore, the difference in electron temperatures between the metals films does not significantly affect the measured thermalization times in this study.

Zarate *et. al.* used a perturbative scattering view of electron thermalization, where the momentum of the initial and final electron states is neglected (the random-k approximation), to formulate $A$ for the free electron gas (FEG) as[6]

$$A = \frac{2\pi}{\hbar} \rho^3 |M_{eff}|^2 \qquad (eq.2)$$

Here, $\rho$ is the density of states (DOS) at the excitation energy, and $M_{eff}$ is an effective screened electron interaction between the excited electron and scattering partner.

The DOS used here was computed using the open-source package Quantum ESPRESSO (Appendix G).[29] The effective screened electron interaction reduces the actual screened electron interaction by assuming that this matrix element does not vary with energy, spin, or momentum.[6,7, 30-34] It has been previously shown that neglecting the energy and momentum dependence is valid below 3 eV.[6,26,27,30] The absorption measurements performed here are not sensitive to electron spin, which justifies averaging the spin components of $M_{eff}$ for Fe and Co.

For Pt, Fe, and Co we use previously determined $M_{eff}$ values.[31,34] To the best of our knowledge, $M_{eff}$ has not been measured or calculated for Mg, however because the DOS for Mg closely aligns with the FEG DOS (Fig. 5), we use the FEG calculation of $M$,[6,36]

$$|M_{FEG}|^2 = \frac{\sqrt{3}\pi}{128} \frac{1}{\rho^3} \frac{\hbar\omega_p}{E_F^2} \tag{eq.3}$$

where $\omega_p$ is the plasma frequency, and $E_F$ is the Fermi level energy.

Table 1 shows the calculated thermalization time for each metal, using equations 1 – 3. More detail on the variables used in these calculations can be found Appendix I. The calculated thermalization times are on the same order of magnitude and follow the experimentally measured trend of $\tau_{Mg} > \tau_{Pt} > \tau_{Fe} > \tau_{Co}$. While the calculated thermalization time for Mg and Pt fall within the error of the experimentally determined thermalization times, the calculated value for Fe is twice as large as the measured value and that calculated for Co is twice as small as the measured value.

Table 1. The thermalization time was calculated for all four metals using equations 1-3. The parameter $\rho$ is equal to the calculated DOS at the central optical pump wavelength of 1.65 eV and the effective screened electron interaction, $M_{eff}$, for Fe and Pt come from Zhukov *et al*, Co comes from Knorren *et. al* and Mg was calculated using eq. 3.[30,33]

|  | Measured therm. time (fs) | $\rho$ (eV⁻¹) | $|M_{eff}|^2$ (eV²) | Calculated therm. time (fs) |
|---|---|---|---|---|
| **Magnesium** | 38 ± 8 | 0.46 | 0.01 | 32 |
| **Platinum** | 15 ± 3 | 0.27 | 0.10 | 17 |
| **Iron** | 4.2 ± 1 | 0.35 | 0.08 | 8.7 |
| **Cobalt** | 2.0 ± 0.3 | 0.34 | 0.64 | 1.1 |

In order to explain the difference between the measured and calculated thermalization times for Fe and Co, the three variables relevant to the calculation of the electron thermalization time, the electron temperature, DOS, and the effective screened electron interaction, are considered. We have already shown that the electron temperature does not significantly affect the results presented here. Previous studies have suggested that the total volume of the DOS participating in scattering principally affects the electron thermalization time, although some of these studies assume the same $M_{eff}$ values for different metals in the process of showing the importance of the DOS.[30,33] The constant volume approximation taken here underestimates the total volume of states participating in scattering, which would overestimate the thermalization time. This could explain the deviation observed in Fe, but not that of Co.

For Co, it is more likely that the assumptions made in determining $M_{eff}$ are responsible for the disparity in the calculated and measured lifetimes. The factors neglected in the effective screened electron interaction are electron spin, energy, momentum, electron localization, and many-body effects, all of which have been attributed to discrepancies between calculated and measured lifetimes and could contribute to either over or underestimating $M_{eff}$ depending on the experimental conditions and specific properties of the metal.[30,35] As the constant DOS assumption does not explain the deviation observed in Co, we believe that $M_{eff}$ is most likely responsible for the calculated electron thermalization time being underestimated for Co. Although we cannot exclude the $M_{eff}$ of Fe contributing to the observed overestimation of the calculated initial electron thermalization lifetime, we favor the simpler explanation that the constant DOS assumption is responsible for the variation observed in Fe. It is beyond the scope of this work to determine which assumptions and excluded factors (i.e. many-body effects, electron spin, electron localization, etc.) are responsible for the differences between the measured and calculated lifetimes. However, this does reflect the importance and difficulty of choosing an appropriate $M_{eff}$, which has already been noted by Zhukov and Chulkov, and it demonstrates the need for further theoretical investigation into these effects, particularly on ferromagnetic metals.[31]

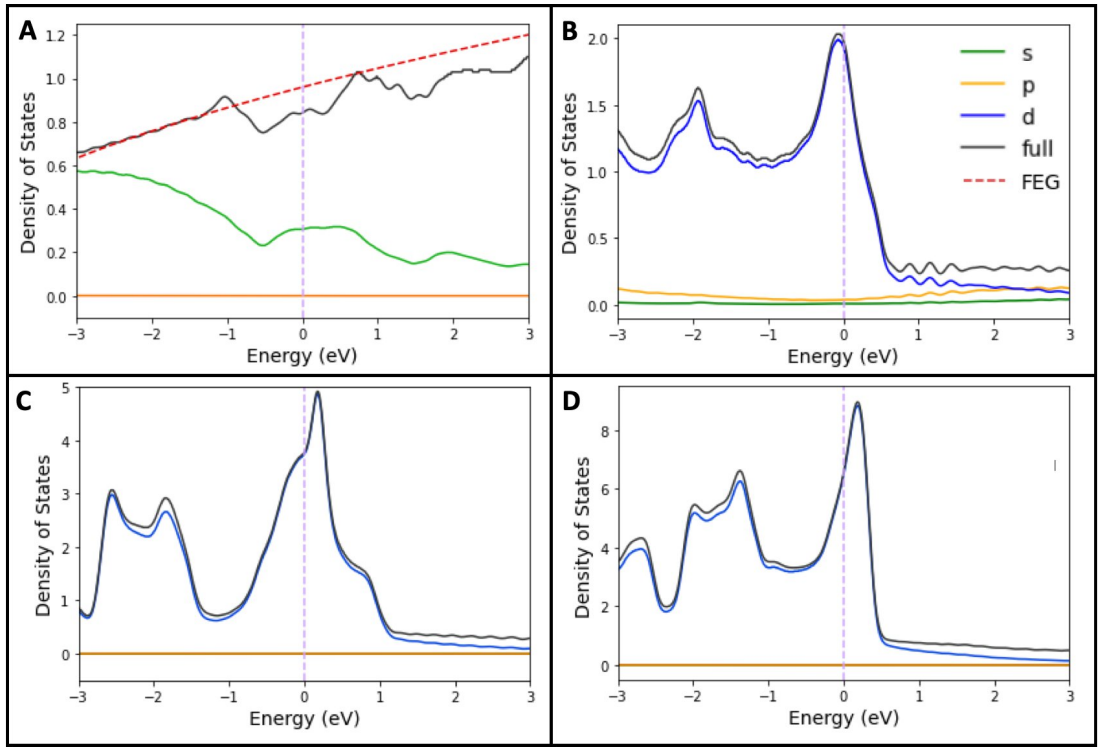

*Fig. 5.* The full and partial density of states for Mg (A), Pt (B), Fe (C), and Co (D) was calculated using DFT and shows the variety of structures of the four metals studied here. The red dashed line shows the free electron gas density of states and the pink dashed line marks the Fermi level position.

Earlier it was observed that Fe and Co both have LFE contributions in their transient XUV absorption lineshape and a faster initial electron thermalization times. This implies that either the LFE influences the initial electron thermalization time or that the basic properties of the metals lead to both a strong LFE and fast initial electron thermalization time. Through the application of the model described above, we get further insight into the correlation between the electron thermalization time and LFE strength. Table 1 shows that the DOS and $M_{eff}$, values are larger as the thermalization time gets faster, with Pt being a notable exception. This suggests that the specific metal properties of the screened electron interaction and the volume of DOS participating in electron scattering are principally responsible for a faster thermalization time, not the LFE. On the other hand, many-body effects, which were shown in section B to have significant influence in Fe and Co, were neglected in the formulation of eq. 1-3. This suggests that the exclusion of the LFE could be responsible for the significant deviations between the experimental and calculated initial thermalization times. Additionally, it is possible that the disparity between calculated and measured thermalization time for Fe and Co potentially comes from the exclusion of many-body effects, such as the LFE, in calculating $M_{eff}$. This implies that the LFE might influence the thermalization time to a much lesser degree than the DOS volume and other factors used in calculating $M_{eff}$. We believe that the most significant contributions to the initial electron thermalization times are the basic metal properties of the DOS volume available for scattering and screened electron interaction and that many-body effects like the LFE contribute less, but still significantly to the initial electron thermalization time.

There are several assumptions made in the formulation of eq. 1-3 that are worth commentary. First, they originate from the view of a single electron excited to a particular energy scattering with a sea of non-excited electrons to reach initial thermalization. The experimental reality contrasts with this simplified view. Experimentally the optical pump excites many electrons ($10^{20}$ – $10^{22}$ electrons cm$^{-3}$) within an energy range dictated by the energy of the broad band optical pump. This results in the equations neglecting the interaction between excited electrons, which we showed contributes to the transient absorption line shape of Co and Fe. It also necessitates the use of an average energy of the ensemble of excited electrons.

Defining an electron temperature for the results also presents a challenge. Directly after optical excitation the excited electrons are in a highly non-equilibrium distribution with no defined temperature. The difficulty in reconciling the experimental reality of no electron temperature with the inclusion of an electron temperature in the equations has led to the commonly used approximation of taking the electron temperature of the hot Fermi-Dirac distribution (formed after initial electron thermalization) as the electron temperature of the non-excited electron bath directly after optical excitation.[9,26,28,37,38] Efforts have been made in the literature to circumvent this known inaccuracy by calculating the full Boltzmann collision integrals for laser excited metals, however this requires in-depth theoretical calculations that are dependent on both the metal properties and experimental conditions, which make this method difficult to generalize.[26] For the calculations presented here we follow the approximation that the electron temperature is equal to the hot Fermi Dirac distribution established after initial electron thermalization. For each metal $k_B T_e$ << E which reduces eq.1 to $\frac{1}{\tau} \approx A\, E^2$. This shows that the difference in electron temperature between the metals does not influence the measured lifetimes.

Despite the application of the single-electron equations to the laser-excited ensemble of electrons and the average values used for the excited electron energy, electron bath temperature, the density of states, and screened electron interaction, the calculated initial thermalization times of Mg and Pt agree well with the experimentally measured initial thermalization times and the general trend of $\tau_{Mg} > \tau_{Pt} > \tau_{Fe} > \tau_{Co}$ is reflected in these calculations. This shows that these simplified, single-electron equations can be used to describe the initial thermalization of an ensemble of laser-excited electrons. Because the quantities used to calculate the initial thermalization times are either basic material quantities

(density of states, plasma frequency, Fermi level energy), readily available from measurements or calculations ($M_{eff}$), and experimentally controlled values (electron bath temperature and excited electron energy), these results can be easily extended beyond the measurements presented in this study, although the limits of the application have not been fully tested in this work.

## II. Conclusion and Outlook

In summary, ATAS was used to determine an initial electron thermalization time of $38 \pm 8$ fs, $15 \pm 3$ fs, $4.2 \pm 1$ fs, and $2.0 \pm 0.3$ fs, for Mg, Pt, Fe, and Co, respectively. TDDFT decomposition of the contributions to the transient absorption lineshape of Mg and Co showed that the LFE is strong in Co, and by extension Fe, and that the LFE is weak in Mg, and by extension Pt. Single-electron equations developed from Fermi-liquid theory within the random-k approximation were used to calculate initial thermalization times of 32 fs, 17 fs, 8.7 fs, and 1.1 fs for Mg, Pt, Fe, and Co, respectively.

There are two main outcomes from this study. First, due to the agreement between the theoretical and measured electron thermalization time and the simplicity of the model used here, this calculation can easily be applied to other metals and potentially to more complicated systems. Second, our results suggest that the initial electron thermalization time of metals is primarily affected by the volume of DOS participating in electron scattering and the screened electron interaction, and that many-body effects, such as the LFE, has a smaller, but significant affect. Ultimately the information gained through this study shows the unique view ATAS contributes to unraveling and monitoring electron dynamics their connection to many body effects in metals and beyond.

## Acknowledgements

Investigations at the University of California, Berkeley were supported by the U. S. Air Force Office of Scientific Research (AFOSR) Grant No. FA9550-19-1-0314 (primary), FA9550-20-1-0334, FA9550-22-1-0451 (DURIP), and the W. M. Keck Foundation No. 046300. Lorenz Drescher acknowledges the European Union's Horizon research and innovation program under the Marie Sklodowska-Curie grant agreement No. 101066334 – SR-XTRS-2DLayMat. Shunsuke A. Sato acknowledges the JSPS KAKENHI Grant Number JP20K14382.

# APPENDIX A: THIN FILM GROWTH

Physical vapor deposition was used to create the metal films. Silicon nitride membranes 30 nm thick (Norcada), which are optically transparent to the XUV probe, were chosen as the deposition substrates. The base pressure in the deposition chamber was ~ $10^{-7}$ Torr. For Pt and Co an ~1 nm/ hr growth rate was achieved by resistively heating a high purity wire (Kurt J. Lesker Company) to 1330 °C – 1348 °C and 1050 °C – 1100 °C, respectively. For Fe and Mg, high purity pellets (Kurt J. Lesker Company) were flash evaporated by a resistively heated sapphire basket that contained the pellet. This method achieved much higher deposition rates of ~250 nm/hr and ~12 nm/hr for Mg and Fe, respectively.

# APPENDIX B: CALCULATING FILM THICKNESS

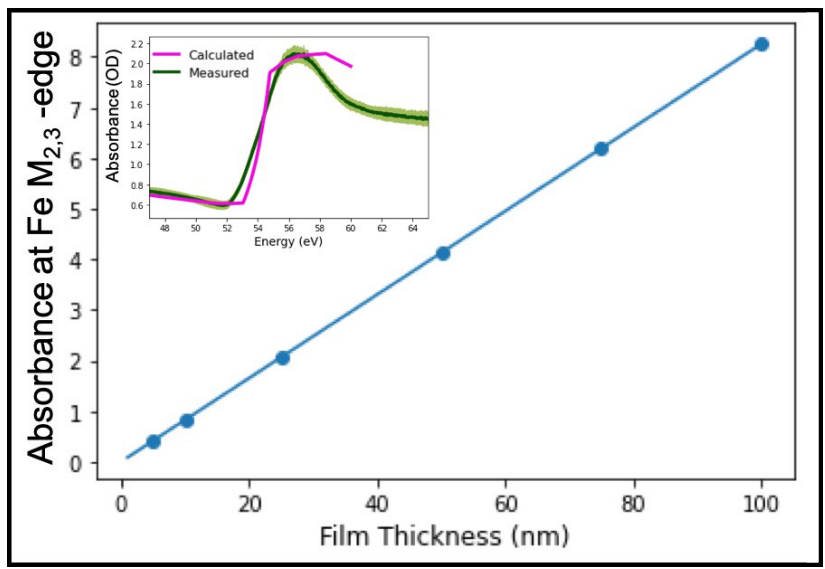

*Fig.6.* The calculated absorption edge of Fe is from the Center for X-ray Optics X-ray solid filter transmission database and was used to determine the thickness of the thermally evaporated Fe film studied here. The insert shows the measured absorption (green line) with shaded area showing the standard deviation and the calculated absorption (pink line).

The XUV absorption measurements without optical pump illumination were used to determine the film thickness of each metal. This is possible because absorbance is directly proportional to the amount of material being probed in the limit of weak absorption. In order to isolate the XUV absorption of the metal film alone, the XUV absorption from a clean 30 nm silicon nitride membrane was subtracted from the XUV absorption of the sample. For Co, Fe, and Mg the Center for X-ray Optics X-ray solid filter transmission database was used to determine the relationship between material thickness and absorbance at the relevant x-ray absorption edge.[39] For example, the transmissions of Fe around the Fe $M_{2,3}$ – edge (52.7 eV) for Fe films of 5 nm, 10 nm, 25 nm, 50 nm, 75 nm and 100 nm were calculated using the Center for X-ray Optics X-ray solid filter transmission data base (Fig. 6).[39] The transmission data were converted to absorbance by taking the negative log of the calculated transmission. A linear regression between the calculated absorbance and film thickness was then used to calculate an Fe film thickness of 18 nm (Fig. 6, inset). The reported variance in film thickness was calculated from the standard deviation of the absorbance measurements using this same method. The same procedure was used for Mg and Co.

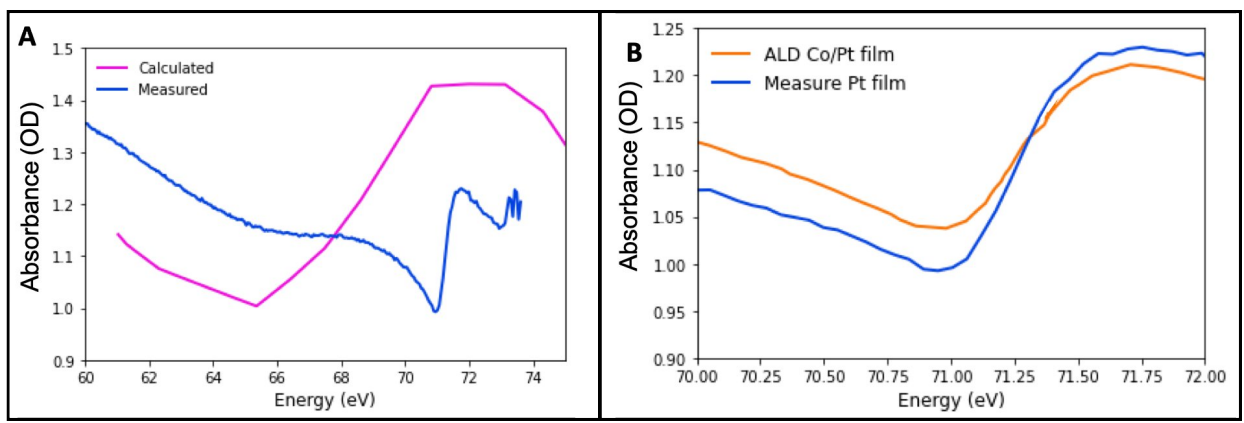

*Fig. 7.* The Fano lineshape of the Pt $N_7$-edge is not well calculated through the Center for X-ray Optics and X-ray solid filter transmission database (A),[39] therefore the absorption from a previously measured Co/Pt film was used to determine the thickness of the Pt film studied here (B). The shaded area denotes the standard deviation in the absorption measurement.

The Fano shape at the Pt $N_7$-edge is not well captured by the calculated spectrum (Fig. 7A). For this reason, a well-calibrated, multilayer Co/Pt thin film was used to calculate the thickness of the Pt film measured in this study (Fig. 7B). The Co/Pt film was grown by atomic layer deposition, where the thickness of each layer was monitored during the growth by reflectometry. There was a total of 21 nm of Pt in the film. By comparing the $N_7$ –edge absorbance for the 21 nm Pt standard to the Pt film studied here, it was determined that the Pt film studied here was 35 nm thick. The variance in film thickness reported in the main text comes from the standard deviation of the absorbance measurement of the Pt film.

## APPENDIX C: TRANSIENT ABSORPTION APPARATUS AND METHODS

Transient absorption studies described here are accomplished through the apparatus outlined in figure 8. A Coherent Legend Elite laser amplifier, which has a central wavelength of 795 nm, a 25 fs (FWHM) temporal duration, a 5 mJ energy, and a 1 kHz repetition rate, is the light source. The laser output is then broadened (Fig. 9A) to 420 nm – 920 nm (Fig. 9B) through self-phase modulation in a stretched, gas-filled, silica hollow core fiber that is 2 m long and has a 704 μm inner diameter. The hollow core fiber is differentially pumped with negligible pressure at the laser entrance and 0.003 psi of Ar at the laser exit of the fiber.

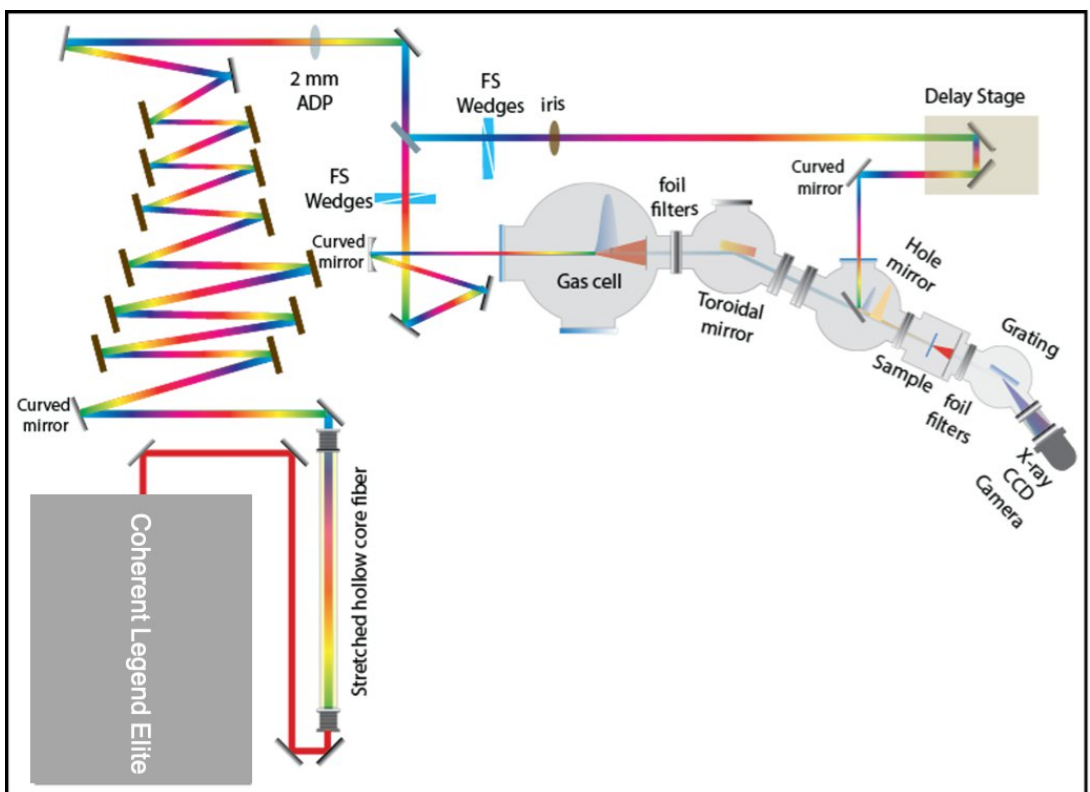

*Fig. 8.* The schematic of the experimental apparatus displays the major components that convert the 795 nm, 25 fs (FWHM) laser output to an attosecond pulse train of extreme ultraviolet light for the absorption probe and a few femtosecond broadband optical pump. It also shows how the optical pump and probe are recombined to perform transient experiments. Here, FS and ADP refer to fused silica and ammonium dihydrogen phosphate, respectively.

The process of self-phase modulation[40] induces a temporal delay between the spectral frequencies of the broadband optical light. Here, double angle chirp mirrors (Ultrafast Innovations PC1332, PC70) are used to correct first order dispersion and 2 mm of ammonium dihydrogen phosphate is used to compensate for higher order dispersion.[41] The beam is then split into the two arms, with 20% of the light reflected to be used as the optical pump, and the remaining 80% to be used in the generation of extreme ultraviolet light absorption probe. Both arms have anti-reflection coated, fused silica wedge pairs, with one wedge on a translation stage, to independently compensate for second order dispersion. After re-compression, the beam typically has a 4 fs pulse duration, as measured through dispersion scanning.

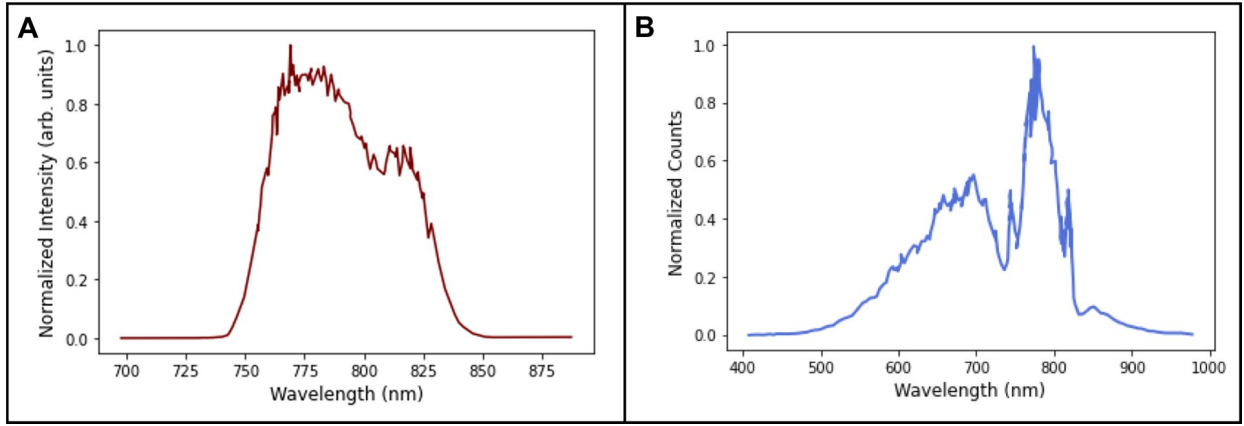

*Fig. 9.* The Coherent Legend Elite output is broadened from ~ 100 nm bandwidth (A) to ~ 400 nm bandwidth (B) through self-phase modulation in a gas-filled stretched hollow core fiber.

The XUV light is generated in the probe arm through the HHG process. This process repeats every half cycle of the 4 fs long, broadband laser field, which inherently imparts an attosecond time duration to each XUV burst. After generation, the residual driving laser field is blocked with a 100 nm thick aluminum foil filter and the diverging XUV light is focused at the sample position with a gold coated toroidal mirror. The transmission of the XUV light through the sample is then measured with an XUV spectrometer, which is constructed from a shallow angle XUV grating and an x-ray sensitized CCD camera. In the pump arm, the optical intensity is controlled with an iris. A double mirror delay stage enables the time delay between the optical pump and XUV absorption probe to be controllably varied. The optical pump is then focused to a spot size of 0.15 mm² at the sample. After the sample, the residual pump light is blocked with a 150 nm thick aluminum foil filter.

Transient absorption spectra are obtained by calculating the change in absorption for each time from the measured XUV transmission in the presence ($I_{pump\ on}$) and absence of the optical pump ($I_{pump\ off}$), $dA = -log \frac{I_{pump\ on}}{I_{pump\ off}}$. This measurement process is then repeated for each specified time delay position in a continuous loop, which enables equivalent time delay points to be averaged. To account for temporal drift, the position on the delay stage corresponding to temporal overlap of the pump and probe is determined in gaseous Ne before each series of time delay measurements is conducted on the metals film.

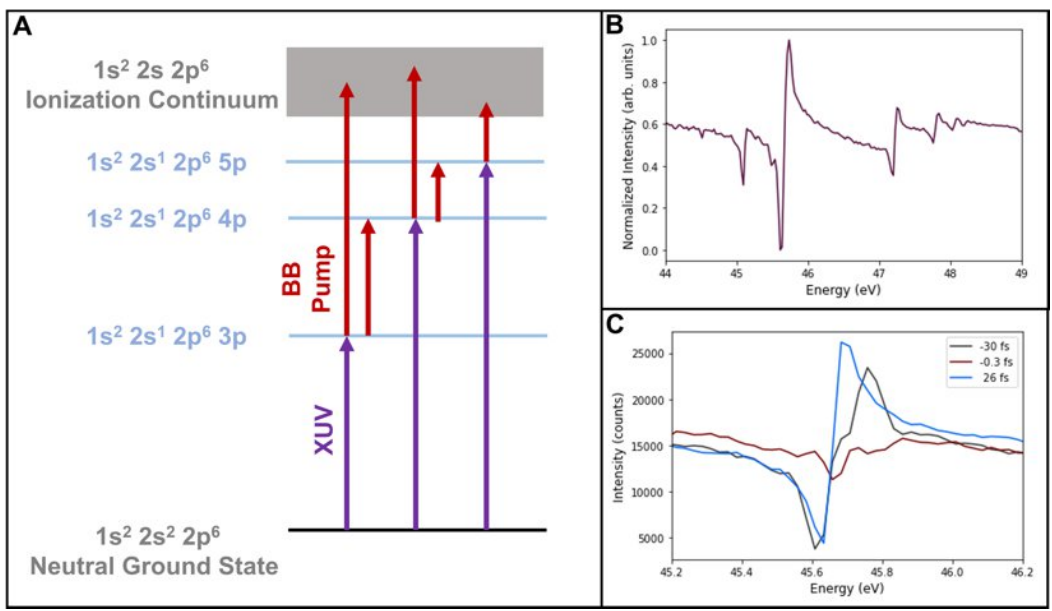

*Fig. 10.* Single electron autoionizing states in Ne couple to the second ionization state to characterize the temporal response function of the instrument. (A) The XUV probe promotes electrons from the ground state into the single electron states. Then the broadband (BB) pump promotes electrons from these single electron states into the ionization continuum. A coupling between the discrete states and the first ionization continuum leads to a Fano profile in the observation of the single electron states with XUV absorption (B). This profile is destroyed at temporal overlap, where the broadband pump excites the population of these single electrons states into the continuum (C).

The temporal duration of the autoionizing one-electron states of Ne is used to account for drifts in temporal overlap throughout the duration of an experiment. XUV absorption of the one electron, *$1s^2 2s^1 2p^6 \, np$* where n = 3, 4, 5..., Ne states can be seen in Fig. 10A. Coupling between the discrete, one electron states and the first ionization continuum, $1s^2 2s^1 2p^6$, results in a Fano profile observed in the XUV absorption (Fig. 10B).[42] At temporal overlap the broadband optical pump excites the one-electron states into the second ionization continuum ($1s 2s^2 2p^6$), destroying the Fano profile (Fig. 10C). As the delay between the optical pump and the XUV absorption probe is increased the Fano line shape is suppressed and then recovers (Fig. 11A). Fitting the line shape response with equation 7 provides the position of temporal overlap (Fig. 11B, dashed orange line) and the time duration of the instrument response, which is equal to the full width half maximum of the fit Gaussian component (Fig. 11B). Fitting the example shown in Fig. 11B shows that temporal overlap is located at 12.1 μm on the delay stage and the instrument response function is 5.2 fs.

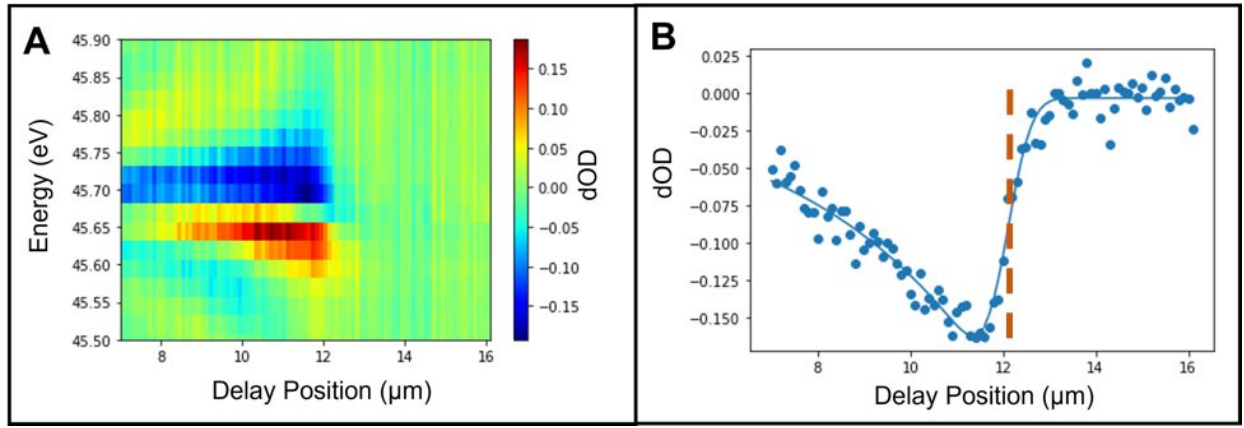

*Fig. 11.* The transient temporal scan for the 1s²2s¹2p⁶ 3p state shows the suppression of the Fano profile around temporal overlap (A). By integrating this spectrum between 45.7 eV – 45.8 eV, blue points, the temporal dependence of this lineshape change can be easily seen (B). Fitting this profile with an exponentially modified error function, solid blue line, shows temporal overlap occurs at 12.1 μm (orange dashed line) on the delay stage and the temporal resolution is 5.2 fs.

Once the delay stage is calibrated, the transmission of the XUV light through the metal sample is collected on the CCD camera at a specified position on the delay stage, first with the optical pump blocked and then with the optical pump on. The XUV transmission through the metal sample with the optical pump blocked and then illuminating the sample is repeated for each specified position of the delay stage. The process of calibrating the delay stage by measuring the lineshape suppression in Ne followed by measuring the XUV transmission through the sample is repeated in order to increase the signal-to-noise ratio of the change of absorption in the sample by averaging the change in absorption at equivalent time delay points. The change in absorbance ($dA$), which is shown in eq. 4, is calculated by taking the negative log of the XUV transmission with optical pump illumination ($I_{pump,\ on}$) divided by the XUV transmission without the optical pump ($I_{pump,\ off}$)

$$DA = -\log \frac{I_{pump,\ on}}{I_{pump,\ off}} \qquad \text{eq. 4}$$

After the change in absorbance is calculated and averaged for each equivalent delay position, after calibration, (Fig. 12, first column) edge referencing is used to account for intensity fluctuations in the XUV source between the recorded XUV transmission with and without optical pump illumination (Fig. 12, second column).[43] The energy ranges used for the reference portion where there is no sample absorption features in the transmitted XUV light are 45 eV – 56 eV for Co, 45 eV – 50 eV for Fe, 58 eV – 68 eV for Pt, and 35 eV – 49 eV for Mg. Lastly, the transient absorption signal appearing before temporal overlap, which occurs from laser heating on the millisecond timescale, is subtracted. In the experiments presented here and in similar work on titanium metal, the temporal dependence of transient absorption features after subtraction was the same as those where the heat signal was not subtracted.[10]

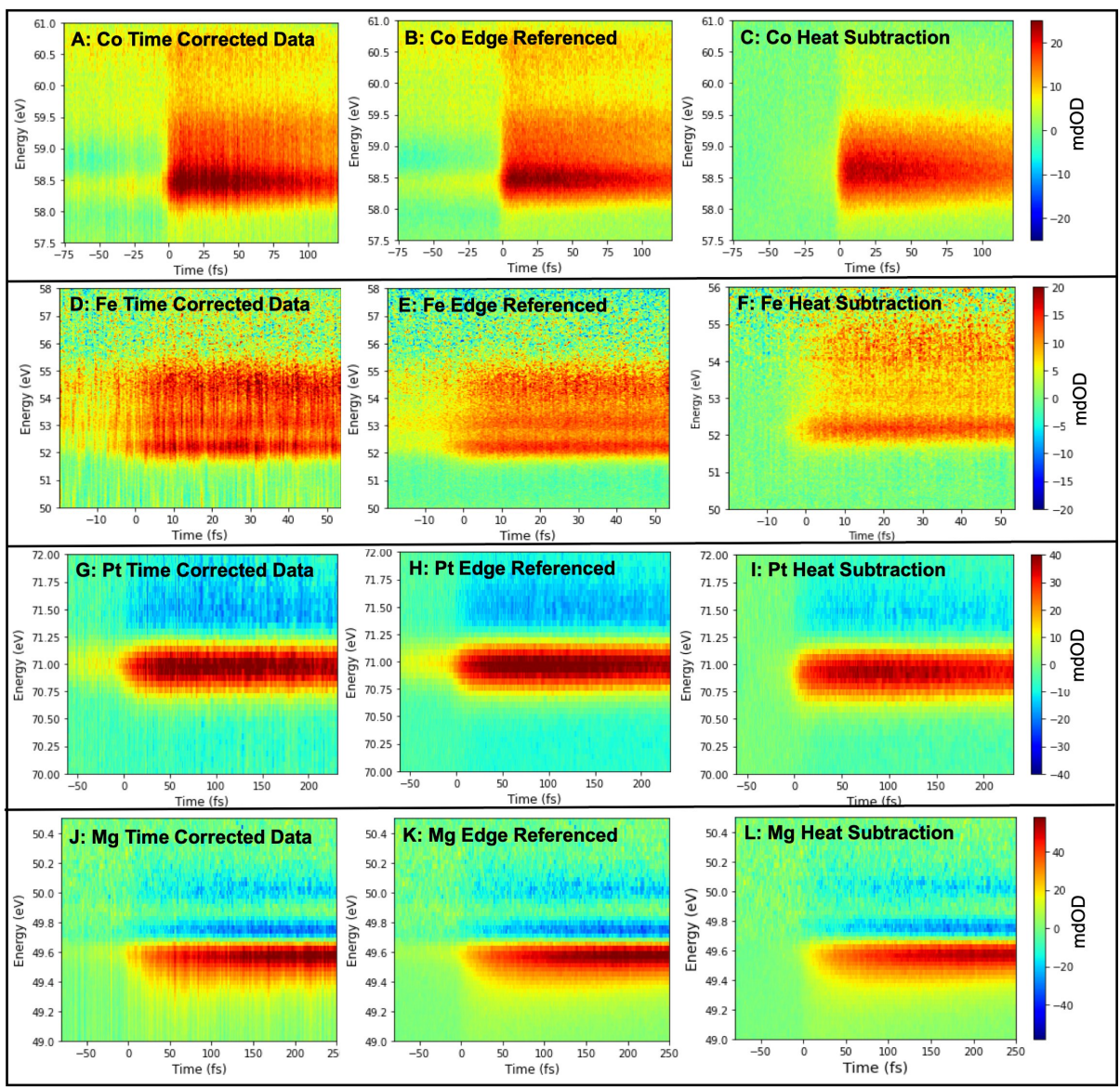

*Fig. 12. The collected transient absorption data for Co (top), Fe (upper middle), Pt (lower middle), and Mg (bottom) with the time axis for each average corrected using the Ne temporal overlap scans are shown in the first column). Next, edge referencing was used to minimize shot-to-shot harmonic fluctuations (middle column). Finally, the latent heat signal, which gives a positive signal before temporal overlap, was subtracted (last column). mdOD is milli delta OD.*

# APPENDIX D: CARRIER DENSITY CALCULAION

Assuming that each photon absorbed by the metal films only excites one electron – hole pair out of the thermal distribution of electrons, the number of excited carriers (N) generated in the metal film is equal to the number of photons in the pump, scaled by how many photons are absorbed by the metal film.[17,18]

$$N = \frac{F}{E\,d}(1-R)\left(1 - e^{-\alpha d}\right)\left(1 + R\,e^{-\alpha d}\right) \qquad \text{eq. 5}$$

Here, F is the pump fluence, E is the energy of the photons, d is the thickness of the film, R is the reflectivity, and $\alpha$ is the absorption coefficient. For each film the extinction coefficient (k) was used to calculate the absorption coefficient ($\alpha$) through the following relationship.

$$\alpha = \frac{4\pi k}{\lambda} \qquad \text{eq. 6}$$

Table 2. A list of the parameters used and the calculated carrier densities.

|  | Thickness (nm) | Extinction Coeff. | Fraction of Reflectivity[44] | Fluence (mJ cm$^{-2}$) | Carrier Dens. (cm$^{-3}$) |
|---|---|---|---|---|---|
| Co | 9.4 | 7.62 | 0.711 | 21.0 | 1.70 x 10$^{22}$ |
| Fe | 18 | 3.59 | 0.605 | 29.2 | 1.88 x 10$^{22}$ |
| Pt | 35 | 7.61 | 0.955 | 22.6 | 1.12 x 10$^{21}$ |
| Mg | 132 | 7.24 | 0.952 | 19.8 | 2.74 x 10$^{20}$ |

While the charge carrier density excited varies for the different metal films this does not significantly affect the electron thermalization times measured in this study. In Appendix H the carrier densities reported here are used to calculate electron temperatures of 1300 K, 2300 K, 3700 K, and 3400 K for Mg, Pt, Fe, and Co, respectively (Appendix H). Equation 1 shows that when $k_B\,T_e \ll E$, then to $\frac{1}{\tau} \approx A\,E^2$. Although electron thermalization generally decreases with increasing electron temperature (carrier density), the variance in electron temperature in the present study does not significantly affect the measured electron thermalization times.

# APPENDIX E: MEASURED TRANSIENT ABSORPTION SPECTRA

The transient absorption spectra for Mg, Pt, Fe, and Co are shown, respectively, in panels A - D of Fig. 13. Edge referencing is used to account for intensity fluctuations in the XUV source between the recorded XUV transmission with and without optical pump illumination (Appendix C Fig. 12).[43] The energy ranges used as the reference portion are 45 eV – 56 eV for

Co, 45 eV – 50 eV for Fe, 58 eV – 68 eV for Pt, and 35 eV – 49 eV for Mg. Lastly, the transient absorption signal appearing before temporal overlap, which occurs from laser heating on the millisecond timescale, is subtracted. In the experiments presented here and in similar work on titanium metal, the temporal dependence of transient absorption features after subtraction was the same as those where the heat signal was not subtracted.[10] The spectra shown in Fig. 13 were integrated over the ranges of 60 – 70 fs, 35 – 45 fs, 15 – 25 fs, and 2 – 12 fs for Mg, Pt, Fe, and Co, respectively, to obtain the transient absorption line shapes shown in Fig. 2.

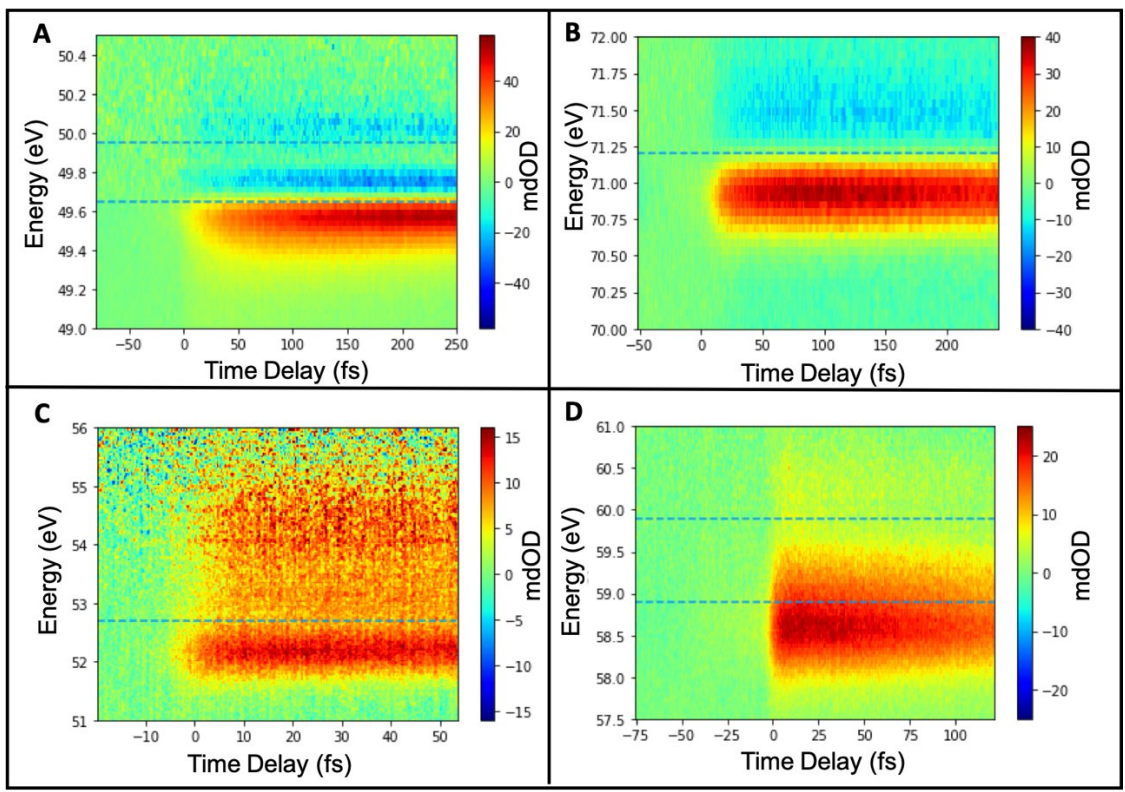

*Fig. 13.* Panels A – D show the transient absorption data collected for Mg (A), Pt (B), Fe (C), and Co (D). The dashed lines correspond to the energy of the static absorption edges. In the color map, a positive change in absorption is displayed in red and a negative change in absorption is displayed as blue.

## APPENDIX F: FITTING TRANSIENT ABSORPTION INCREASE OVER TIME

The positive feature shown in Fig. 2 A - D was integrated with respect to energy in order to assess how this feature changes over time. The integration windows used were 49.2 eV – 49.7 eV, 70.5 eV – 71.3 eV, 51.8 eV - 52.6 eV, and 56.5 eV – 65.2 eV for Mg, Pt, Fe, and Co,

respectively. These data were then fit with an exponentially modified Gaussian function of the form:

$$f(x,\mu,\sigma,\lambda) = \int_{-\infty}^{a} \frac{\lambda}{2} exp\left[\frac{\lambda}{2}\left(2\mu + \lambda\sigma^2 - 2x\right)\right] erfc\left(\frac{\mu + \lambda\sigma^2 - x}{\sqrt{2}\,\sigma}\right) dx \qquad \text{eq. 7}$$

Here, $\mu$ is the mean of the Gaussian component, $\lambda$ is the rate of the rise, and $\sigma$ is the variance of the Gaussian component. In the context of the transient XUV absorption experiments performed here, $\mu$ corresponds to the time where both the optical pump and XUV absorption probe arrive simultaneously, $\sigma$ is the temporal instrument response function, which is measured in Ne as described below, $\lambda$ is the rate of the rise of the signal due to the electron thermalization, and the integration bound ($a$) represents the time points. The timescales reported in the main text are the inverse of the rate of rise.

## APPENDIX G: TIME DEPENDENT DENSITY FUNCTIONAL THEORY AND DENSITY FUNCTIONAL THEORY

Here, we describe the numerical methods for analyzing static electronic structure and electron dynamics from first-principles. In order to compute the electronic structures of metals, we employed an open-source package, Quantum ESPRESSO, which is based on density functional theory.[29] For practical calculations, we used the projector augmented-wave method to treat valence electrons, and we utilized the PBE functional as the exchange-correlation functional.[45,46] We set the cutoff energy to 45 Ry for all metals investigated in this work. The partial density of states (projected density of states) was computed by projecting the computed Kohn-Sham orbitals onto atomic orbitals.

In order to describe electron dynamics in metals, we employed time-dependent density functional theory (TDDFT).[47] For laser-excited metals following pump irradiation, the absorption spectra were computed using the finite-electron temperature method, which models hot-electron states.[48] The alteration in absorption spectra due to an increase in electron temperature was further analyzed by decomposing it into the state-filling effect and the local field effect, employing the Fermi-Dirac distribution.[12]

Here, we briefly describe the theoretical methods for analyzing the optical properties of laser-excited solids. The details of the methods were discussed elsewhere.[49] The dynamics of electrons are described by solving the following time-dependent Kohn-Sham equation for each electronic orbital:

$$i\hbar \frac{\partial}{\partial t} u_{bk}(r,t) = \left[ \frac{1}{2m_e}(p + \hbar k + eA(t))^2 + \hat{v}_{ion} + v_{HXC}(r,t) \right] u_{bk}(r,t) \qquad \text{eq. 8}$$

where $b$ is the band index, and $k$ is the Bloch wavevector, $A(t)$ is a spatially-uniform vector potential, $\hat{v}_{ion}$ is the ionic potential, and $v_{HXC}(r,t)$ is the Hartree-exchange-correlation potential. In this work, we employ the local density approximation together with the adiabatic approximation for describing the electron dynamics in metals, and the Hartree-exchange-correlation potential becomes thus a functional of the instantaneous electron density as[50]

$$v_{Hxc}(r,t) = v_{Hxc}[\rho(r,t)](r,t) \qquad \text{eq. 9}$$

Here, the electron density is defined by

$$\rho(r,t) = \sum_b \int_{BZ} dk f_{bk}^{T_e} |u_{bk}(r,t)|^2 \qquad \text{eq. 10}$$

with the occupation factor $f_{bk}^{T_e}$ given by the Fermi-Dirac distribution. Hence, the occupation factor depends on the electron temperature $T_e$.

By employing the time-dependent orbital, $u_{bk}(r,t)$, the time-dependent current density can be computed as

$$J(t) = -\frac{e}{m_e \Omega} \sum_b \int_{BZ} dk f_{bk}^{T_e} \int_\Omega dr\, u_{bk}^*(r,t) v_k(t) u_{bk}(r,t) \qquad \text{eq. 11}$$

where $\Omega$ is the volume of the unit cell, and $v_k(t)$ is the velocity operator.

In this work, to compute the absorption spectra of metals, we compute the electron dynamics by solving eq. 8 with a weak impulsive distortion, $A(t) = -E_0 e_p \Theta(t)$, with the field amplitude of $E_0$ and the polarization direction along $e_p$. We further compute the electric current with eq. 11 for evaluating the optical conductivity $\sigma(\omega)$ as

$$\sigma(\omega) = \frac{\int_0^\infty dt e^{i\omega t - \gamma t} e_p \cdot J(t)}{E_0} \qquad \text{eq. 12}$$

where $\gamma$ is a damping factor introduced to reduce numerical noise due to the finite simulation time. In this work, we set $\gamma$ to 0.5 eV/$\hbar$. By using the optical conductivity, one can further evaluate the dielectric function and the absorption coefficient as

$$\epsilon(\omega) = 1 + \frac{4\pi i}{\omega}\sigma(\omega) \qquad \text{eq. 13}$$

$$\mu(\omega) = \frac{2\omega}{c}\Im\left[\sqrt{\epsilon(\omega)}\right] \qquad \text{eq. 14}$$

For practical calculations for the optical properties of metals, we employed an open-source package, Octopus.[50] For bulk magnesium, we employ a hexagonal crystal structure, and we set the lattice parameter $a$ to 3.21 Å and the lattice constant ratio $c/a$ to 1.624.[51] The hexagonal lattice is discretized into $24 \times 24 \times 39$ real-space grid points, and the first Brillouin zone is discretized into $16^3$ $k$-points. The magnesium atom is described using a norm-conserving pseudopotential method, treating 2s, 2p, and 3s electrons as valence.[52] Likewise, for bulk cobalt, we use the hexagonal crystal structure, and we set $a$ to 2.51 and $c/a$ to 1.611.[51] The lattice is discretized into $24 \times 24 \times 38$ grid points, and the first Brillouin zone is discretized into $24^3$ $k$-points. The cobalt atom is described using a norm-conserving pseudopotential method, treating 3s, 3p, 3d and 4s electrons as valence.[52]

We compute the absorption spectra of Mg and Co in the equilibrium phase by setting the electron temperature $T_e$ to 300 K, while we compute those of laser-excited systems by setting $T_e$ to 0.25 eV/$k_B T$ and 0.026 eV/$k_B T$ for Mg and Co, respectively. Furthermore, the change in the absorption spectra by the increase in the electron temperature were decomposed into three different components: One is the state-filling effect, which originates from the temperature dependence of the Fermi-Dirac distribution via the occupation factor in eq. 11. Another one is the change of the local-field effect, which originates from the temperature dependence of the occupation factor in the time-dependent part of the potential in eq. 9. The other is the band structure modification, which originates from the temperature dependence of the static part of the potential in eq. 9. Note that, the band structure modification is negligible for both Mg and Co, in this work.

## APPENDIX H: CALCULATING FINAL ELECTRON TEMPERATURE

The electron temperature reached after initial thermalization ($T_{th}$) can be estimated by the following relationship.

$$\frac{hc}{\lambda} N = \int_{T=300\,K}^{T_{th}} C_e(T')\, dT' \qquad \text{eq. 15}$$

Here, N is the number of photons absorbed by the sample (Appendix E), $\lambda$ is the wavelength of the light, and $C_e(T)$ is the electron heat capacity. The left hand side is the energy absorbed by the sample and the right hand side is the integrated electron heat capacity.

For each metal, the electron heat capacity was estimated by using the linear relationship between heat capacity and temperature with the slope taken as the Somerfield electron heat capacity coefficient.[53-55] The final electron temperature was found to be 1300 K, 2300 K, 3700 K, and 3400 K for Mg, Pt, Fe, and Co, respectively (Appendix H). We assess the impact that the difference in electron temperature between the metals has on the initial electron thermalization time through eq. 1. For all the metals, $k_B T_e \ll E$, which reduces eq. 1 to $\frac{1}{\tau} \approx A\, E^2$. The result shows that the temperature differences between the metals does not significantly affect the measured thermalization time.

## APPENDIX I: $\rho$ AND $M_{eff}$ USED IN THE CALCULATED ELECTRON THERMALIZATION TIME

Table 1 presents the DOS ($\rho$) and screened electron interaction ($|M_{eff}|^2$) values used in equations 1 and 2 to calculate the electron thermalization time. Within the random-k approximation theory for calculating thermalization times there have been several ways of including the number of states participating in scattering. The most accurate of these accounts for the volume of states with excited electrons by integrating up to the excitation energy of the unfilled states and accounts for the scattering partners by integrating the filled DOS up to the Fermi level.[6,31,32] While this is more accurate, we choose the more accessible simplification outlined by Zhukov and Chulkov where a constant DOS is assumed, which reduces the scattering integrals to cubing a singular DOS value at the excitation energy.[31] The DOS used here was calculated with an open-source package, Quantum ESPRESSO, with the methodology outlined in Appendix G. For Mg and Co there are two atoms per unit cell, therefore the DOS values at 1.65 eV are divided by 2, in order to get the per atom values.[32]

Although it is possible to calculate $M_{eff}$ within the free electron gas (FEG) model, for most metals the FEG approximation is innacurate.[31,32] However, it has been shown that fitting photoemission data with the random-k approximation or directly calculating it with many-body scattering theory (GW, GW+T, STA-DOS) provide a more accurate $M_{eff}$ value.[7,31,33,35] For this reason, we use the $M_{eff}$ values for Fe and Pt provided by Zhukov *et. al* and the $M_{eff}$ value reported for Co by Knorren *et. al*.[30,33] As the $M_{eff}$ value for Mg has not been reported, we used the FEG approximation shown in equation 3 to calculate $M_{eff}$ for Mg. While this is not accurate for most metals, because the DOS of Mg is similar to the FEG DOS, Mg is a good candidate for this simplification. The plasma frequency and Fermi level were calculated using the free electron model.[56] In this model $\omega_p = 47.1\ eV \left(\frac{r_s}{a_0}\right)^{\frac{-3}{2}}$, where $\frac{r_s}{a_0}$ is the reduced electron density parameter. For Mg, $\frac{r_s}{a_0}$ is 2.66 which gives $\omega_p$ = 16.5 eV. For the Fermi level, $E_F = \frac{\hbar^2}{2m_e} (3\ \pi^2\ n)^{\frac{3}{2}}$ where $m_e$ is the mass of an electron and *n* is electronic density. For Mg, n = 8.61 x $10^{22}$ cm$^{-3}$, yielding $E_F$ = 7.08 eV. [56]